\documentclass[pre,superscriptaddress,twocolumn,showkeys,showpacs]{revtex4}
\usepackage{epsfig}
\usepackage{latexsym}
\usepackage{amsmath}
\begin {document}
\title {Slow dynamics in a driven two-lane particle system}
\author{Adam Lipowski}
\affiliation{Faculty of Physics, Adam Mickiewicz University, 61-614
Pozna\'{n}, Poland}
\author{Dorota Lipowska}
\affiliation{Institute of Linguistics, Adam Mickiewicz University,
60-371 Pozna\'{n}, Poland}
\begin {abstract}
We study a two-lane model of two-species of particles that perform biased diffusion. Extensive numerical simulations show that when bias $q$ is strong enough oppositely  drifting  particles form some clusters that block each other. Coarsening of such clusters  is very slow and their size increases logarithmically in time. For smaller $q$ particles collapse essentially on a single cluster whose size seems to diverge at a certain value of $q=q_c$.
    Simulations show that despite slow coarsening, the model has rather large power-law cooling-rate effects. It makes its dynamics different from glassy systems, but similar to some three-dimensional Ising-type models (gonihedric models).
\end{abstract}
\maketitle
Despite its apparent simplicity, statistical mechanics of low-dimensional driven diffusive particle systems is still attracting a considerable interest~\cite{ZIADOMB}. This is due to numerous applications ranging from vehicular or pedestrian transport~\cite{SANTEN}, to gel electrophoresis~\cite{ALON}, to molecular motors~\cite{HOWARD}. Moreover, such systems exhibit the wealth of highly nontrivial and often surprising  features. Indeed, in the steady state of these systems we observe novel universality classes~\cite{JANSSEN}, off-critical long-range correlations~\cite{GARRIDO}, or internal energy that is a decreasing function of temperature~\cite{ZIA2002}. It shows that experience we have gained studying equilibrium systems is not very helpful in the realm of nonequilibrium phenomena.

There is a growing evidence that dynamics of these nonequilibrium systems is also very interesting. For example,
asymptotic cluster growth in a certain two-species model is governed by the exponent that is twice as large as could be
(naively) predicted using Lifshitz-Slyosov theory~\cite{GEORGIEV06}. Nucleation of clusters in this model is also surprising: residence distribution develops a peak that signals formation of a macroscopic cluster, but that peak disappears, however, when the system size becomes extremely large~\cite{GEORGIEV05}.

 For some other models Evans et al. predicted~\cite{KAFRI} that power-law coarsening, that typically accompanies symmetry breaking transitions, should become logarithmmically slow. However, due to numerical difficulties such a behaviour was observed only in an effective toy model. 
Such a slow coarsening is one of the important ingredients of glassy dynamics. Absence of quenched disorder  means that kinetic barriers which slow down the evolution are generated by the dynamics of the model. A similar scenario takes place in the so-called ordinary glasses~\cite{RITORT}. Another important ingredient of glassy dynamics are very small cooling-rate effects, i.e., very slow (presumably logarithmically slow) growth of the characteristic length scale as a function of the inverse of the cooling rate. Some three-dimensional Ising-type models exhibit such a behaviour~\cite{LIPDES}. Although generation of barriers (even diverging ones) might lead to slow-coarsening it does not necessarily imply small cooling-rate effects~\cite{LIP2000}. In addition, some arguments suggest that in two-dimensional Ising-type models thermal fluctuations should suppress generation of energy barriers restoring fast, power-law coarsening~\cite{SHORE1992}. The role of dimensionality in the glassy dynamics is important also in off-lattice systems. Indeed, recent Molecular Dynamics simulations of one-, {two-,} three-, and four-dimensional Lennard-Jones systems show that the tendency to form glasses increases with dimensionality and in one-dimensional systems glassy transition was absent~\cite{KOB2009}. Let us notice, however, that simulations of off-lattice systems are computationally very demanding and it is often mere presence or absence of hysteresis that is used to identify the glassy transition.

The main objective of the present paper is to check whether absence of glassy transition in one-dimensional systems might be a property also of lattice models. Slowly coarsening driven diffusive particle systems are good candidates for such an analysis. To examine whether other ingredients of glassy dynamics might appear in such low-dimensional systems it would be desirable to analyse appropriately defined cooling rate effects. Since analytically tractable driven diffusive models usually can be solved only with respect to steady state properties~\cite{BLYTHE}, further understanding of dynamics of these systems most likely have to rely on numerical approaches. In systems with slow dynamics development of efficient computational methods is particularly important. In the present paper, we examine a model where two species of particles perform a biased diffusion on a one-dimensional (two-lane) lattice.
Closely related models have already been analysed in the context of formation of spatial structures~\cite{ZIA92},  nucleation kinetics, and cluster growth~\cite{GEORGIEV05,GEORGIEV06} and our work contributes to better understanding of this class of models.
Our simulations with an efficient algorithm clearly indicate that the regime with spatial inhomogeneities is characterised by logarithmically slow coarsening. However, rather large, power-law cooling-rate effects  show that phase transition separating homogeneous and inhomogeneous phases of the model differs from realistic glassy transition. We also discuss similarities of the dynamics of our model  and that of some Ising-type models.

\begin{figure}
\vspace{0.0cm} \centerline{ \hspace{2cm}\epsfxsize=11cm
\epsfbox{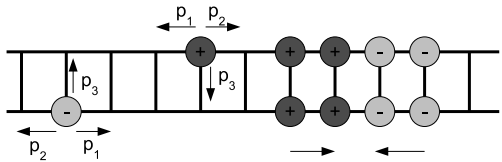} } \vspace{-5cm} \caption{Probabilities of elementary transitions in our model. For $p_2>p_1$ positive/negative particles are preferentially moving to the right/left and some gridlocks might form as the one to the right. However, collective movements of particles might create some holes and eventually eliminate such gridlocks. Since such movements are against the drive and thus are highly unlikely, the lifetime of such gridlocks rapidly increases with their size.}
\label{epsilon}
\end{figure}

In our model there are $N$  particles of two kinds, "positive" and "negative", that perform biased diffusion on a two-lane lattice of size 2x$L$ (all results reported in this paper are obtained for density $\frac{N}{2L}
=0.1$). Each site of the lattice can be occupied by at most one particle, and we assume that the number of "positive" and "negative" particles are equal. In addition, particles are exposed to "electric" field that introduces some bias on their diffusion  in the horizontal direction. Detailed dynamics is specified below:\\
\indent (i) Select randomly a particle.\\
\indent (ii) Select the target site (one of the three nearest neighbours of the selected particle) with probability depending on the kind of the selected particle: For "positive" particles left and right neighbours are selected with probabilities $p_1$ and $p_2$, respectively; For "negative" particles these probabilities are exchanged (see Fig.\ref{epsilon}). The probability of inter-lane jumps $p_3$ is for both kinds of particles the same and all results described in the present paper are obtained for $p_3=0.5$. An important parameter of the model, that is related with the strength of the "electric" field, is the bias probability $q$ defined using the difference of probabilities of right and left jumps as $q=\frac{1}{2}(p_2-p_1)$, and of course, $p_1+p_2+p_3=1$. Let us also notice that dynamics preserves CP symmetry (i.e., probability of jumping to the right for a "positive" particle is the same as that of jumping to the left for a "negative" particle).\\
\indent (iii) Provided that the target site is empty, the selected particle jumps to that site.Otherwise, the attempt to move the selected particle is rejected.\\

Closely related models have already been analysed by Schmittmann {\it et al.}~\cite{ZIA92,GEORGIEV05,GEORGIEV06}. In their formulation of the dynamical rules, however, it is a lattice bond that is selected randomly. As a result, their algorithm is efficient but only for dense systems (when density is small there is a large probability of selecting a bond with no particles at its ends and no moves are possible). In our case we can ensure that each time we select a particle. However, when large clusters of particles are formed, a lot of particles get blocked. In such a case it is useful to keep the list of active particles (i.e., those that can move), and select particles only from such a list~\cite{ODOR}. We have to notice, that since a unit of time in our model is defined as a single, on average, update of each particle, selecting only from the list of active particles the simulation time $t$ is increased by $1/N_A$, where $N_A$ is the (current) number of active particles. 
Our algorithm was actually a hybrid version: when the number of active particles is large, we select a particle out of all particles, and only when it drops below the threshold we switch to the  version where we select out of the list of active particles (as a threshold we used the value $N_A=N/5$). Let us also notice that in the late-stage evolution there is usually a small number of active particles and the described algorithm is much more efficient than that where selection is made out of all particles.

\begin{figure}
\vspace{0.5cm} \centerline{ \hspace{0cm}\epsfxsize=9cm
\epsfbox{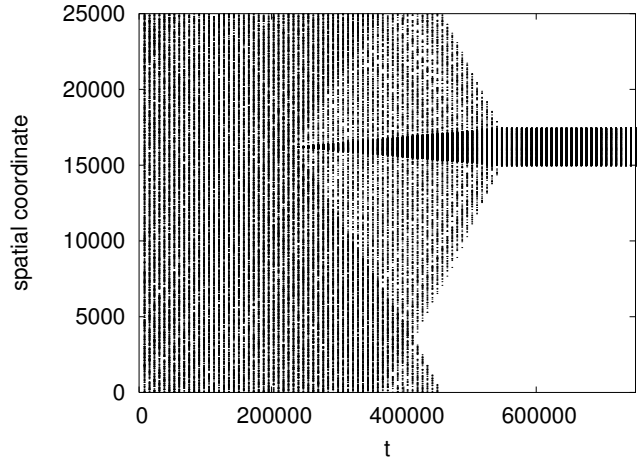} } \vspace{0cm} \caption{The time
evolution of a random initial configuration of $N=5\cdot 10^3$
particles on a ladder of length $L=25\cdot 10^3$ and for
$q=0.045$. After around $t=5\cdot 10^5$ MC steps essentially all particles form a single cluster. The configuration of every fifth particle is recorded every 7500 MC steps.} \label{config0045}
\end{figure}

It is already known~\cite{ZIA92} that for models of this kind for sufficiently small but positive bias $q$ the model remains in the homogeneous phase with a slow drift of particles ("positive" to the right and "negative" to the left). For larger $q$ it was reported~\cite{ZIA92} that after some time a large cluster of particles is formed and the drift stops.
Formation of such a cluster is illustrated in Fig.~\ref{config0045} that shows the time dependence of the initially random configuration of $5\cdot10^3$ particles evolving at $q=0.045$. Indeed, one can notice that around $t=2.5\cdot10^5$ a cluster is formed that grow and after $t\sim 5\cdot10^5$ almost all particles collapse on that cluster.

\begin{figure}
\vspace{0.5cm} \centerline{ \hspace{0cm}\epsfxsize=9cm
\epsfbox{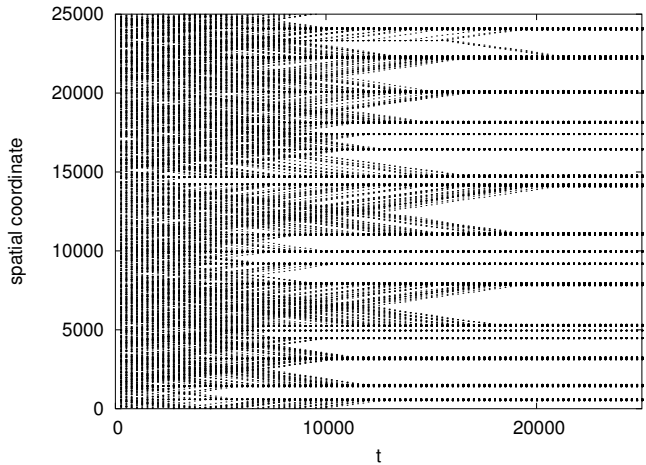} } \vspace{0cm} \caption{The time
evolution of a random initial configuration of $N=5\cdot 10^3$
particles on a ladder of length $L=25\cdot 10^3$ and for
$q=0.1$. The configuration of every fifth particle is recorded every 250 MC steps.} \label{config01}
\end{figure}

Main results reported in the present paper are inspired by the observation that for larger bias $q$ instead of forming a single cluster, the model gets trapped in the slowly-coarsening state with the wide spectrum of clusters of various sizes. An example of such an evolution is shown in Fig.~\ref{config01}. 

\begin{figure}
\vspace{0.5cm} \centerline{ \hspace{0cm}\epsfxsize=9cm
\epsfbox{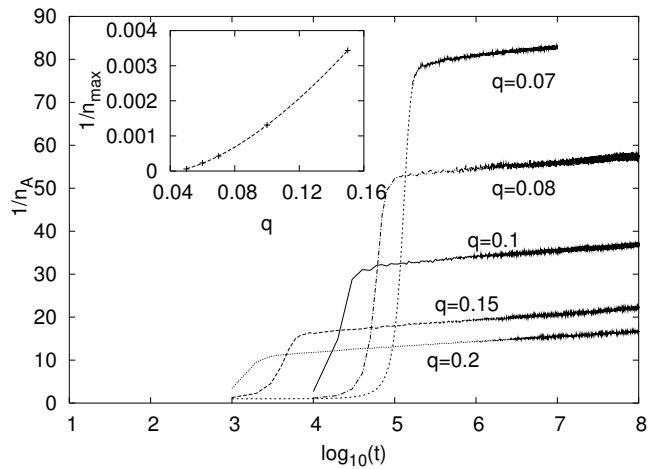}
}
\vspace{0cm} \caption{The time evolution of the inverse of the fraction of active particles $n_A$ (semi-log scale). Simulations were made for $N=2\cdot10^4$ and averaged over $10^2$ independent runs. The inset shows the $q$ dependence of the inverse of the maximum cluster size $N_{max}$. These data suggest that $N_{max}$ diverges around $q=q_c\sim 0.041$.}
\label{quench}
\end{figure}

To examine in more detail the coarsening we calculated the average fraction of active particles $n_A=N_A/N$ as a function of time for several values of $q$ (Fig.~\ref{quench}). Since active particles are mainly at the border of clusters~\cite{COMMENT1}, their number $N_A$ is proportional to the number of clusters  and the characteristic cluster size should scale as $N/N_A=1/n_A$. On the semi-log plot of our data one can clearly see that $1/n_A$ and thus the characteristic cluster size, after an initial fast growth, at late stage increases logarithmically slowly in time ($\sim {\rm log}(t)$). We also measured the average size of the maximum cluster $N_{max}$ just after the initial fast growth. Our data (inset in Fig.~\ref{quench}) suggest that  this quantity diverges around $q=q_c\sim 0.041$.

\begin{figure}
\vspace{0.5cm} \centerline{ \hspace{0cm}\epsfxsize=9cm
\epsfbox{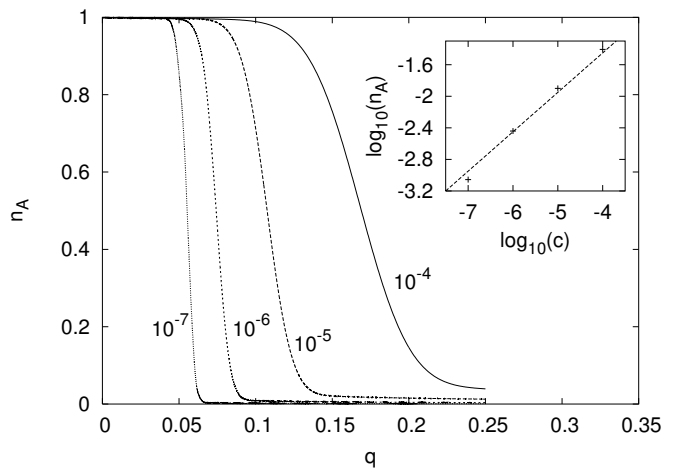}
}
\vspace{0cm} \caption{The fraction of active particles $n_A$ as a function of $q$ during the linear cooling with the values of cooling rate $c$ indicated. Results are averages over $10^2$-$10^4$ independent runs ($N=2\cdot 10^5$).
The inset shows the value of $n_A$ at the end of the cooling ($q=0.25$) as a function of cooling rate $c$ (log-log scale). The dotted line has a slope 0.5 that corresponds to $n_A\sim c^{0.5}$.}
\label{cool}
\end{figure}

Logarithmically slow coarsening suggests that the dynamics of the model might exhibit some glassy features. Considering small- and large-$q$ phases as analogs of high- and low-temperature phases, respectively, one can examine evolution of our model under continuous cooling. In our simulations we increase $q$ linearly in time from 0 till 0.25 as $q(t)=ct$, where $c$ can be interpreted as a cooling rate. The average value of $n_A$ as a function of $q$ is shown in Fig.~\ref{cool}. One can notice that for slow cooling $n_A$ rapidly decreases around $q=0.04$ and such a value agrees with the divergence of maximum cluster size $N_{max}$ (inset in Fig.~\ref{quench}). In glassy dynamics it is of interest to examine the zero-temperature characteristic length $l_0$ and its dependence on the cooling rate. Some arguments~\cite{SHORE1992} and Monte Carlo simulations~\cite{GREST} suggest that in glassy systems  $l_0$ should only very weakly increase with  the decreasing cooling rate. Very small dependence on the cooling rate is also reported in some Molecular Dynamics simulations~\cite{KOB} as well as in some experiments~\cite{BRUNING}. Linear fit to the average value of $n_A$, as measured at the end of the cooling, as a function of the cooling rate $c$ suggests that $n_A(q=0.25)\sim c^{0.5}$ (inset in Fig.~\ref{cool}). Such a relation would imply that the characteristic cluster size at the end of the cooling increases as $c^{-0.5}$.

Taking into account that the model has logarithmically-slow coarsening, such a fast increase of the characteristic length  seems to be rather surprising. In our opinion, there are two possible scenarios that might explain such a behaviour. First, there might be a regime close to the transition point $q_c<q<q_c'$ where the coarsening dynamics is much faster (with presumably power-law coarsening). The dominant part of the coarsening would take place during the time spent in that regime and power-law increase of the characteristic length would be the expected feature. As shown by Shore et al., similar scenario takes place in a certain three-dimensional Ising model with next-nearest neighbour  interactions~\cite{SHORE1992}. Our results on coarsening in Fig.~\ref{quench} extends only up to $q=0.07$ and we cannot exclude that for smaller $q$ a power-law coarsening would operate (i.e., $q_c'$, if exists, should be smaller than 0.07).
Another possibility is that the regime with logarithmically slow coarsening extends up to the transition point $q=q_c$ but the dynamics at this point being sufficiently fast to generate such a growth. Such a situation most likely occurs in three-dimensional gonihedric Ising models and is caused by vanishing of energy barriers precisely at the critical temperature~\cite{LIP2000}.

Although the dynamics of our model has much larger cooling-rate effects than that expected in glasses, we cannot exclude that some other one-dimensional system will be more realistic with this respect. Let us notice that in three-dimensional Ising model with 4-spin (plaquette) interaction, which is a particular version of the already mentioned gonihedric model, there are very small, and presumably logarithmic, cooling-rate effects (as well as logarithmically slow coarsening)~\cite{LIPDES}.

In conclusions, using extensive numerical simulations we have shown that in the two-lane two-species driven diffusive particle model formation of clusters proceeds via logarithmically slow coarsening. Adopting phenomenology developed for glassy systems we implemented a cooling protocol for our model. Obtained results  show that the model misses an important ingredient of glassy dynamics and despite slow coarsening it has rather strong, power-law cooling-rate effects. Together with some Molecular Dynamics simulations of off-lattice systems~\cite{KOB}, our work suggests that absence of glassy transition might be a more general feature of one-dimensional systems. Of course, one cannot exclude that a certain modification of our model with e.g., different density of particles, different concentrations of "positive" and "negative" particles, or jumping rules without CP symmetry will behave differently, exhibiting both slow coarsening and small cooling-rate effects. Alternatively, there might be some more fundamental reasons prohibiting glassy dynamics in one-dimensional systems, and
 further research that would resolve that problem, in our opinion, should be undertaken.

\textbf{Acknowledgments:} This research was supported with grant
N~N202~071435. We gratefully acknowledge access to the computing
facilities at Pozna\'n Supercomputing and Networking Center.
 

\begin{thebibliography}{}
\bibitem{ZIADOMB} For reviews, see e.g., B.~Schmittmann and R.~K.~P.~Zia, {\it Statistical Mechanics of Driven Diffusive Systems} in Phase Transitions and Critical Phenomena Vol.~17, edited by C.~Domb and J.~L.~Lebowitz (Academic Press, New York 1995).
\bibitem{SANTEN} D.~Chowdhury, L.~Santen, and A.~Schadschneider, Phys.~Rep.~\textbf{329}, 199 (2000).
\bibitem{ALON} U.~Alon and D.~Mukamel, Phys.~Rev.~E \textbf{55}, 1783 (1997).
\bibitem{HOWARD} J.~Howard, Nature (London), \textbf{389}, 561 (1997).
\bibitem{JANSSEN} H.~K.~Janssen and B.~Schmittmann, Z.~Phys.~B \textbf{64}, 503 (1986).
\bibitem{GARRIDO} P.~L.~Garrido, J.~L.~Lebowitz, C.~Maes, and H.~Spohn, Phys.~Rev.~A \textbf{42}, 1954 (1990).
G.~Grinstein, D.~H.~Lee, S.~Sachdev, Phys.~Rev.~Lett.~\textbf{64}, 1927 (1990).
\bibitem{ZIA2002} R.~K.~P.~Zia, E.~L.~Praestgaard, and O.~G.~Mouritsen, Am.~J.~Phys.~\textbf{70}, 384 (2002).
\bibitem{GEORGIEV06} I.~T.~Georgiev, B.~Schmittmann, and R.~K.~P.~Zia, J.~Phys.~A \textbf{39}, 3495 (2006).
\bibitem{GEORGIEV05} I.~T.~Georgiev, B.~Schmittmann, and R.~K.~P.~Zia, Phys.~Rev.~Lett.~\textbf{94}, 115701 (2005).
\bibitem{KAFRI} M.~R.~Evans, Y.~Kafri, H.~M.~Koduvely,  and D.~Mukamel, Phys.~Rev.~E {\bf 58}, 2764 (1998). Y.~Kafri, D.~Biron, M.~R.~Evans and D.~Mukamel,
Euro.~Phys.~J. B {\bf 16}, 669 (2000).
\bibitem{RITORT} F.~Ritort and P.~Sollich, Adv.~in Phys.~{\bf 52}, 219 (2003).
\bibitem{LIPDES} A.~Lipowski and D.~Johnston, Phys.~Rev.~\textbf{61}, 6375 (2000).
\bibitem{LIP2000} A.~Lipowski, D.~Johnston, and D.~Espriu, Phys.~Rev.~\textbf{62}, 3404 (2000).
\bibitem{SHORE1992} J.~D.~Shore, M.~Holzer, and J.~P.~Sethna, Phys.~Rev.~B \textbf{46}, 11376 (1992).
\bibitem{KOB2009} R.~Br\"uning, D.~A.~St-Onge, S.~Patterson, and W.~Kob, J.~Phys.: Cond.~Matt.~{\bf 21}, 035117 (2009).
\bibitem{BLYTHE} R.~A.~Blythe and M.~R.~Evans, J.~Phys.~A {\bf 40}, R333 (2007).
\bibitem{ZIA92} B.~Schmittmann, K.~Hwang, and R.~K.~P.~Zia, Europhys.~Lett.~\textbf{19}, 19 (1992).
\bibitem{ODOR} Such algorithms are frequently used, for example, in studying models with absorbing states (for review see G.~\'Odor, Rev.~Mod.~Phys.~{\bf 76}, 663 (2004); H.~Hinrichsen, Adv.~Phys.~{\bf 49}, 815 (2000)). One can consider them as a zero-temperature limit of the continuous-time algorithm (A.~B.~Bortz, M.~H.~ Kalos, and J.~L.~Lebowitz, J.~Comput.~Phys. {\bf 17}, 10 (1975)).
\bibitem{COMMENT1} Some active particles might be located inside clusters but during coarsening clusters compactify and we expect that such configurations are very rare especially at late stage. 
\bibitem{GREST} G.~S.~Grest, C.~M.~Soukoulis, and K.~Levin, Phys.~Rev.~Lett.~{\bf 56}, 1148 (1986).
\bibitem{KOB} K.~Vollmayr, W.~Kob and K.~Binder, Europhys.~Lett.~{\bf 32}, 715 (1995).
\bibitem{BRUNING} R.~Br\"uning, and M.~Sutton, Phys.~Rev.~B {\bf 49}, 3124 (1994).
\end{thebibliography}
\end {document}